\documentclass[11pt,twoside]{article}
\usepackage{asp2010}

\resetcounters

\begin{document}
\bibliographystyle{asp2010}

\title{Reorienting Our Perspective of Broad Absorption Line Quasars}
\author{Michael A. DiPompeo$^1$, Michael S. Brotherton$^1$, Carlos De Breuck$^2$, Sally Laurent-Muehleisen$^3$
\affil{$^1$Dept. of Physics and Astronomy 3905, University of Wyoming, 1000 E. University, Laramie, WY 82071, USA}
\affil{$^2$European Southern Observatory, Karl Schwarzschild Strasse 2, 85748 Garching bei M\"{u}nchen, Germany}
\affil{$^3$Illinois Institute of Technology, 3101 South Dearborn St., Chicago, IL 60616, USA}}

\begin{abstract}
New multi-frequency radio observations of a large sample of radio-selected BAL quasars, along with a very well matched sample of normal quasars, are presented.  The observations were made one immediately after the other at 4.9 and 8.4 GHz with the goal of measuring the radio spectral index of each source.  We have identified, for the first time, a significant difference in the spectral index distributions of BAL versus non-BAL quasars, with BAL sources showing an overabundance of steep-spectrum sources.  This is the first direct observation suggesting that BAL quasars are more likely to be seen farther from the radio jet axis, although a range of orientations is needed to explain the width of the distribution.  Utilizing a few different relationships between spectral index and viewing angle, we have also performed Monte-Carlo simulations to quantify the viewing angle to these sources.  We find that the difference in the distributions of spectral index can be explained by allowing the BAL sources to have viewing angles extending about 10 degrees farther from the jet axis than non-BAL quasars.
\end{abstract}

\section{Introduction}
Broad absorption lines (BALs) are present in around 20\% of optically selected quasar spectra, and the intrinsic fraction is likely much higher \citep{Knigge2008}.  The blueshift and widths of these lines suggest high-velocity outflows along the line of sight, reaching to several percent of the speed of light.  Although these features have been recognized for decades, it is still an open question as to why only some quasars show the signature of these outflows.  

Over the years, a dichotomy has developed in attempting to explain the BAL subclass.  One common explanation is orientation, where BAL clouds are lifted from the surface of the accretion disk and radiatively driven into an equatorial outflow with a covering fraction of 10-20\% \citep{Elvis2000}.  Optical polarization properties have been argued to support this picture \citep{Ogle1999}, as well as the similarities in the emission lines of BAL and non-BAL sources \citep{Weymann1991}.  While this model is successful in explaining some observations of BAL quasars, it has its shortcomings, and has never been shown directly to be correct.

Radio properties are useful diagnostics for orientation.  Resolved jets and lobes can allow direct observation of source orientation, but the majority of BAL quasars are point-like at the arcsecond resolutions typical of big surveys like FIRST.  Because about 50\% of normal quasars show extended structure compared to about 10\% for BALs \citep{Becker2000}, a second explanation began to emerge- that BAL quasars were an early phase in the lifetime of all quasars \citep{Gregg2002, Gregg2006}.  For non-extended sources, the radio spectral index $\alpha$ ($f \propto \nu^{\alpha}$, where $f$ is the radio flux and $\nu$ is the frequency) is an indicator of orientation, at least statistically for a sample.  As a radio jet points more along the line of sight (``face-on''), relativistic beaming boosts the core component, which is optically thick and has a flat radio spectrum.  Farther from the jet axis (``edge-on''), the radio lobes dominate the flux, and being optically thin they have a steeper spectrum.  There is scatter to this relationship, but a sample seen only edge-on compared to a sample seen mostly face-on should show a clear difference in spectral index distributions.  This has been searched for, and so far (and in small samples) no difference in spectral index distributions between BAL and non-BAL quasars has been found \citep{Becker2000,Montenegro2008}.

To do this properly, larger samples are needed- this work finally lays to rest the question of whether BAL and non-BAL quasars have different spectral index distributions, and therefore different orientations.

\section{Targets \& Observations}
The sample was built by cross-matching the \citet{Gibson2009} catalog of BAL quasars from SDSS DR5 and the FIRST survey.  Only sources with FIRST fluxes above 10 mJy were included, and a redshift cut of 1.5 was applied to ensure inclusion of the \ion{C}{IV} emission line in the spectrum.  The final sample includes 74 BAL quasars.  In order to make meaningful comparisons, a sample of 74 individually matched (in redshift, SDSS $i$-band magnitude, and FIRST flux) unabsorbed quasars was also developed.

Observations were made over two periods with the VLA/EVLA at 4.9 and 8.4 GHz (1.4 GHz measurements are already available from FIRST).  With the exception of a few cases, measurements at both frequencies were performed within 30 to 60 minutes of each other, to remove any complications in measuring radio spectral index due to radio variability.  Data from other surveys was also collected when available, ranging in frequency from 15 GHz down to 74 MHz.  In most cases however, only our new data and the FIRST measurements are available.

\section{Measurements, Modeling, and Results}
\
Using the new fluxes, we measured the radio spectral indices of all sources in several different ways.  First, we utilized two-point spectral indices between 4.9 and 8.4 GHz ($\alpha_{8.4}^{4.9}$; with simultaneous fluxes) and 1.4 and 4.9 GHz ($\alpha_{4.9}^{1.4}$; with non-simultaneous fluxes).  We also applied a simple linear fit (assuming a power-law spectrum) to all available data points gathered from the literature combined with our new values, using the slope as $\alpha_{fit}$.  We then compared the distributions of of these indices for the BAL and non-BAL samples, using both Kolmogorov-Smirnov (K-S) tests and Wilcoxon Rank-Sum (R-S) tests.  The distributions of $\alpha_{8.4}^{4.9}$ for the two samples are shown in Figure 1.  The statistical results ($D_{KS}$ for the K-S test and $Z_{RS}$ for the R-S tests, along with corresponding $P$ values) are shown in Table 1; the top half of the table shows the results including all sources, and the bottom half shows the results restricting the samples to only compact sources to eliminate any effects of resolved structure.  We see that regardless of the test performed or restriction to compact sources, the samples are significantly different at a 3 to 4 $\sigma$ level.  This suggests that BAL quasars do show a preference for steeper spectra and edge-on orientations, although the widths of the distributions also show that both samples cover a range of orientations.

\begin{table}[!t] 
\caption{Statistical tests on $\alpha$ distributions.} 
\smallskip 
\begin{center} 
{\small 
\begin{tabular}{ccccccc} 
\tableline 
\noalign{\smallskip} 
Measurement & $n$ BAL & $n$ non-BAL & $D_{ks}$ & $P_{ks}$ & $Z_{rs}$ & $P_{rs}$\\
\noalign{\smallskip} 
\tableline 
\noalign{\smallskip} 
$\alpha_{8.4}^{4.9}$         & 72 & 72 & 0.347 & 0.0002 & 4.00 & $3.1\times 10^{-5}$ \\[3pt]
$\alpha_{4.9}^{1.4}$         & 73 & 73 & 0.287 & 0.0036 & 3.18 & 0.0007                        \\[3pt]
$\alpha_{fit}$                      & 73 & 74 & 0.322 & 0.0007 & 3.76 & $8.4 \times 10^{-5}$ \\[3pt]
\hline \\
\hline \\
c $\alpha_{8.4}^{4.9}$      & 63 & 56 & 0.337 & 0.0016 & 3.63 & 0.0001                        \\[3pt]
c $\alpha_{4.9}^{1.4}$      & 63 & 57 & 0.342 & 0.0012 & 3.70 & 0.00011                      \\[3pt]
c $\alpha_{fit}$		         & 63 & 58 & 0.394 & 0.0001 & 4.19 & $1.4 \times 10^{-5}$ \\[3pt]
\noalign{\smallskip} 
\tableline
\end{tabular} 
} 
\end{center} 
\end{table} 

\begin{figure}[!t]
\caption{$\alpha_{8.4}^{4.9}$ comparison for BAL and non-BAL quasars.}
%\plotfiddle{f5.eps}{vsf=70}{hsf=70}
\plotone{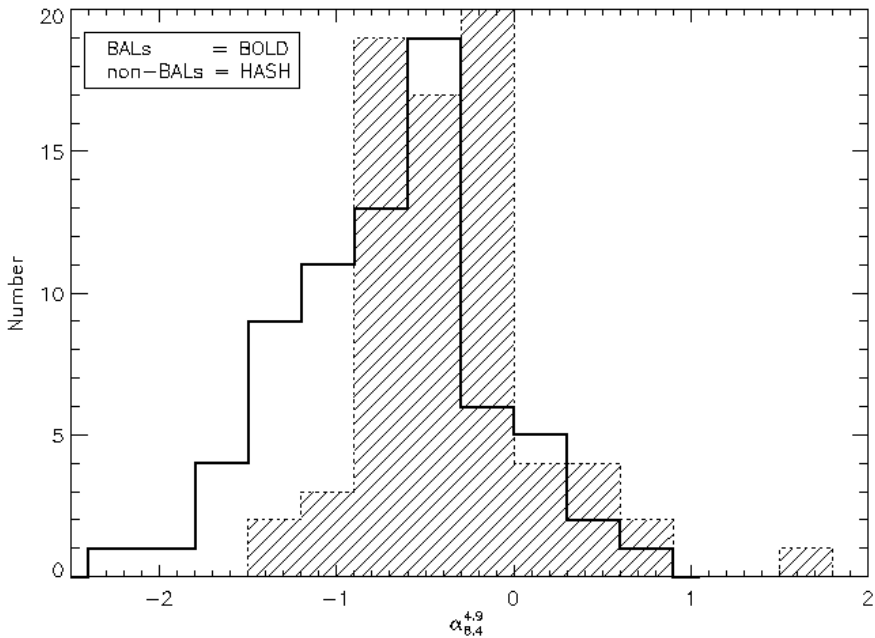}
\end{figure}

In order to quantify this result, we have done some Monte-Carlo modeling of the $\alpha$ distributions.  This requires use of a model relationship between $\alpha$ and viewing angle ($\theta$).  We do this in two ways, one based purely on observation and one from semi-empirical simulations.  The observational model uses the sample of \citet{Wills1995}, for which values of $\theta$ were obtained via superluminal motion observations.  After collecting radio fluxes from NED to find spectral indices, we were able to do a linear fit to the relationship.  To simulate the scatter in this relationship, we used the standard deviation of the distribution of $\alpha$ for the quasars in the 3CRR catalog.  Because this catalog is based on low frequency observations, it is most likely almost entirely lobe dominated sources and all the variation is probably due to intrinsic differences between sources and not because of any beaming effects.  The semi-empirical relationship is based on the simulations of \citet{Wilman2008}, and the scatter there is determined directly from the data.  The overall shape of the the two relationships are similar, though the scatter in the observational model is larger than that from the empirical simulations.

We then simulate random bi-polar jets in 3-D space, calculate the viewing angle, and assign a spectral index based on one of the above models.  We do this 74 times to build a simulated sample of the same size as the real one, and then compare the $\alpha$ distributions to the actual data.  We repeat this process $10^5$ times, and measure the probability of reproducing the observed spectral index distributions.  We can then restrict the allowed viewing angles to the sources and run the experiment again, in order to determine which range of viewing angles most often produces a match to the observed data.  Viewing angles above 45 degrees are not allowed, as it is assumed that beyond that line of sight obscuration from dust is likely to become an issue; e.g. \citet{Barthel1989}.

\begin{figure}[!b]
\caption{Simulation results using the semi-empirical model; BALs are on the left, unabsorbed quasars on the right.}
\plottwo{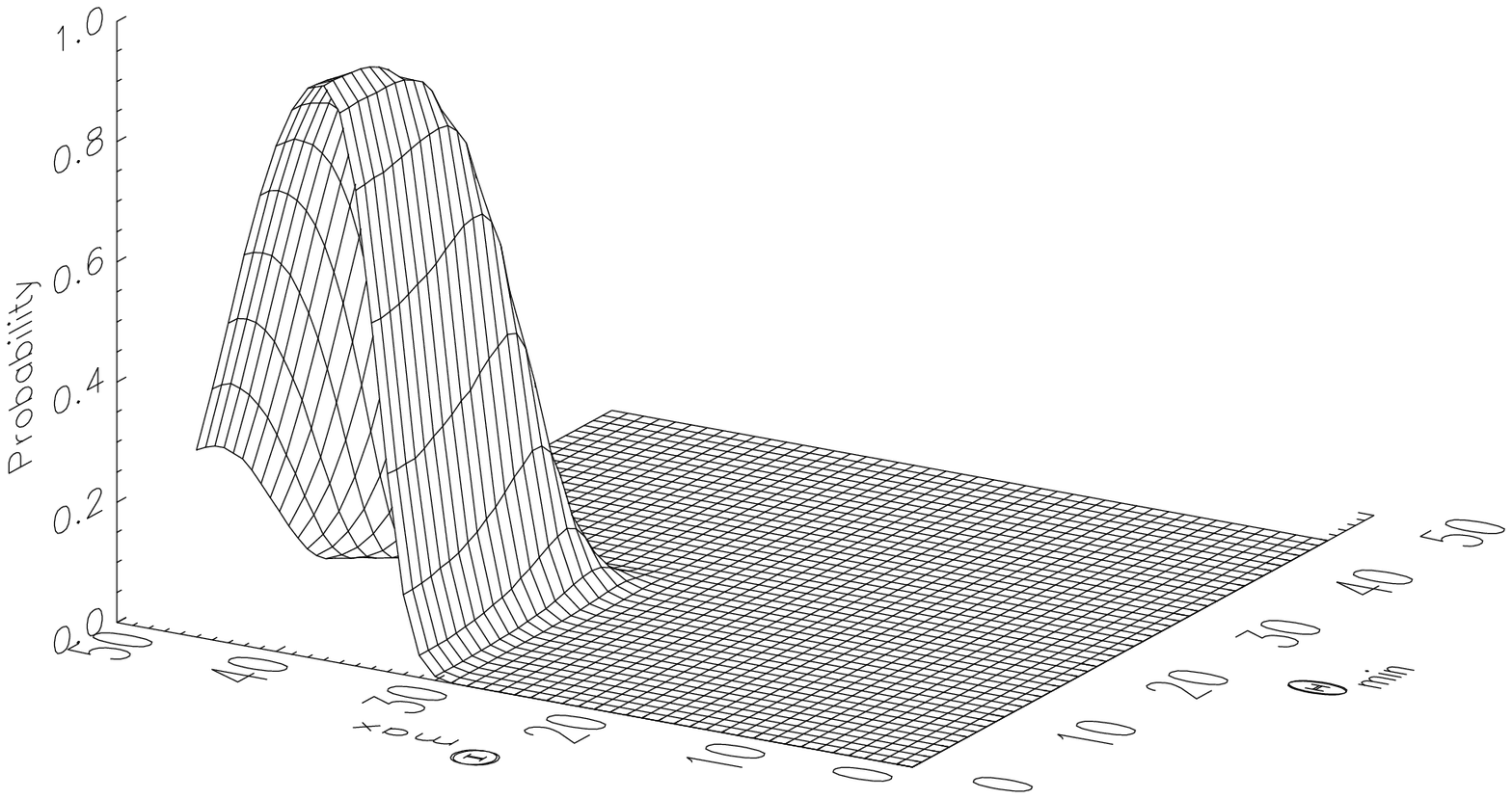}{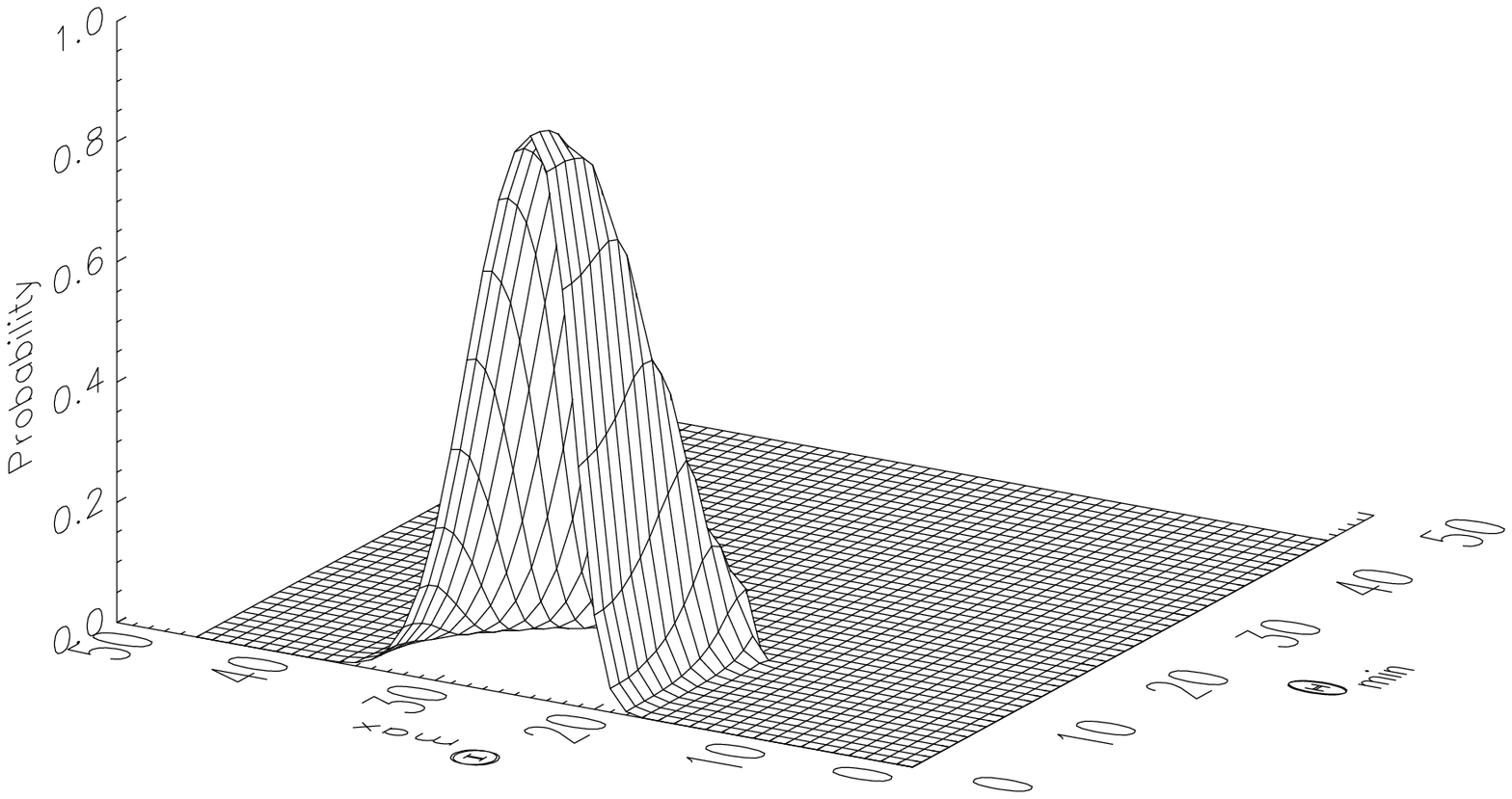}
\end{figure}

The results when comparing the observed distributions of $\alpha_{8.4}^{4.9}$ to the simulations based on the semi-empirical relationship between $\alpha$ and $\theta$ are shown in Figure 2 (the left side is the result for BAL quasars, the right side is for non-BALs).  The x-axis is $\theta_{min}$ (the lowest allowed viewing angle), the y-axis is $\theta_{max}$ (the largest allowed viewing angle), and the z-axis is the probability of each viewing angle range reproducing the observed result (assuming a value of $P_{KS}>0.05$ indicates that the distributions are from the same parent population).  The most probable viewing angle range for the BAL sample is from 1 to 37 degrees, compared to 0 to 24 degrees for the non-BAL sample.  So while both samples are seen all the way down to 0 degrees (along the jet axis), the BAL sample extends about 10 degrees more edge-on compared to non-BALs.  Running the simulations with the observationally determined model the results are similar, except that the probability is not as obviously peaked, but flatter in the $\theta_{min}$ direction due to the larger scatter in the $\alpha$-$\theta$ relationship.  However, the range of viewing angles covered with a probability of matching observations of greater than 90\% is almost identical to what was found with the empirical model- BAL sources extend to viewing angles about 10 degrees more edge-on.  

This is the first direct observation to indicate that orientation does indeed play a role in the presence of BALs, and they are seen out to larger viewing angles compared to non-BAL sources.  However, the results also show that orientation is not likely to be the only factor, as the models clearly show they are seen along many of the same lines of sight as unabsorbed quasars.  The orientation versus evolution dichotomy is likely a false one, and we need to consider the role of both in order to fully understand these objects.

\acknowledgements We would like to acknowledge the Wyoming NASA Space Grant Consortium for funding a portion of this work and providing travel support to attend this meeting.  We would also like to thank the European Southern Observatory for awarding DGDF funding to M. DiPompeo to visit and collaborate with C. De Breuck, which was instrumental in this work.

\bibliography{author}

\end{document}